\documentclass[10pt, twocolumn]{iopart}
\usepackage{scicite}
\usepackage{times}

\newcounter{lastnote}

\usepackage{color}

\usepackage{graphicx}

\begin{document}



\title{A gravitational wave observatory operating beyond the quantum shot-noise limit: Squeezed light in application}


\author{J.~Abadie$^{1}$,	
B.~P.~Abbott$^{1}$,	
R.~Abbott$^{1}$,
T.~D.~Abbott$^{59}$    	
M.~Abernathy$^{2}$,	
C.~Adams$^{3}$,	
R.~Adhikari$^{1}$,	
C.~Affeldt$^{4,11}$,	
P.~Ajith$^{1}$,
B.~Allen$^{4,5,11}$
% ...
}	
\address{$^{1}$LIGO - California Institute of Technology, Pasadena, CA  91125, USA }
\address{$^{2}$SUPA, University of Glasgow, Glasgow, G12 8QQ, United Kingdom }
\address{$^{3}$LIGO - Livingston Observatory, Livingston, LA  70754, USA }
\address{$^{4}$Albert-Einstein-Institut, Max-Planck-Institut f\"ur Gravitationsphysik, D-30167 Hannover, Germany}
\address{$^{5}$University of Wisconsin--Milwaukee, Milwaukee, WI  53201, USA }
% ...
\address{$^{11}$Leibniz Universit\"at Hannover, D-30167 Hannover, Germany }
% ...
\address{$^{59}$California State University Fullerton, Fullerton CA 92831, USA}
% ...






\vspace{0.6cm} 
\hspace{1.8cm} Corresponding author:  ~R.~Schnabel$^{4,11}$, email: roman.schnabel@aei.mpg.de\\

\vspace{1.2cm}

\noindent 
\textbf{Around the globe several observatories are seeking the first direct detection of gravitational waves (GWs). These waves are predicted by Einstein's General Theory of Relativity \cite{Ein16} and are generated 
 e.g. by black-hole binary systems
~\cite{SSc09}
. Current GW detectors are  Michelson-type 
kilometer-scale laser interferometers {measuring the distance changes between in vacuum suspended mirrors}. 
The sensitivity of these detectors at frequencies above several hundred Hertz is limited by 
the vacuum (zero-point) fluctuations of the electromagnetic field. A quantum technology -- the injection of squeezed 
light~\cite{Cav81} -- offers 
a solution to this problem. 
Here we demonstrate the squeezed-light enhancement of GEO\,600, which will be the GW observatory operated by the LIGO Scientific Collaboration in its search for GWs for the next 3-4 years. 
GEO\,600 now operates with its best ever sensitivity which proves 
the usefulness of quantum entanglement and 
the qualification of squeezed light as a key technology {for future} GW astronomy.\\[5mm]
}

\indent Just as electromagnetic radiation is produced by accelerated charges, gravitational waves are generated by accelerated mass distributions, such as supernova explosions or neutron star and black hole binary systems spiraling into each other~\cite{SSc09}. GWs propagate at the speed of light and reveal themselves as an alternating stretching and compressing of space-time, transverse to their direction of propagation. The direct measurement of GWs is extremely challenging. Two neutron stars merging at the other side of our galaxy would produce a maximum space strain amplitude $h$ here at the Earth of just $h\approx10^{-19}$.
The strain would be just $h\approx 10^{-22}$ in the case that the same event is located somewhere in the Virgo cluster of galaxies being about 1000 times farther away. 
Michelson-type laser interferometers are suitable observatories to measure gravitational waves \cite{Wei72}.
%

Currently a global network comprising 2 LIGO observatories in the USA (with an arm length of 4 km~\cite{AbbottETAL09ligo}, not operational since Nov. 2010 due to upgrade activity), the Virgo project of the European Gravitational Observatory (with 3\,km arms~\cite{Acernese08}) and the German-British detector GEO\,600 (with 600\,m arms~\cite{Wetal02,Gretal10}) exists, with further observatories planned or proposed in Japan~\cite{AraiETAL09}, Australia~\cite{BarrigaETAL10} and Europe~\cite{PunturoETAL10}. The targeted GW-frequency band extends from below 10\,Hz to about a few kHz. An ideal probe for space-time disturbances requires test masses which are free-falling in the direction of the laser beams. Therefore, the interferometer mirrors are suspended as sophisticated multistage pendulums and are situated in ultra-high-vacuum systems.

\begin{figure*}[t!]
\centerline{\includegraphics[width=\linewidth]{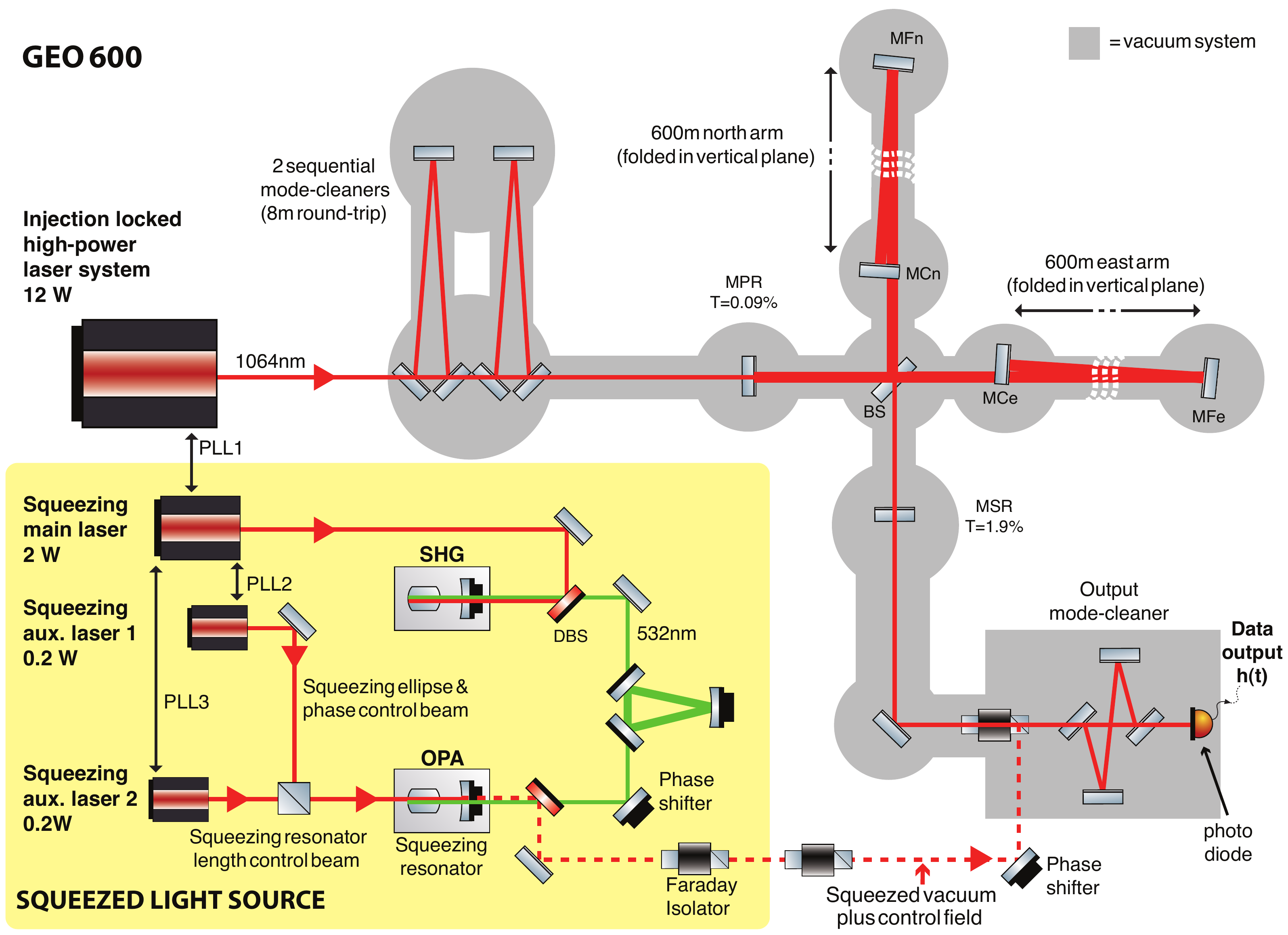}}
  \vspace{0mm}
\caption{A simplified optical layout of the squeezed-light enhanced gravitational wave {observatory} GEO\,600, which consists of the conventional GEO\,600 and the additional squeezed light source (yellow box, {see method summary for details). The observatory has two singly folded arms with a total optical length of 2400\,m. A GW passing from most directions will shorten one arm, while the length of the perpendicularly orientated arm is increased, and vice versa in the next half-cycle of a passing wave, producing a periodic power change of the  output light that is detected by a photo diode. The observatory is operated such that almost all the light is back-reflected towards the 12\,W input laser system, by keeping the interferometer output on a dark fringe by a control system. A power-recycling mirror (MPR) leads to a resonant enhancement of the circulating light power of 2.7\,kW at the beamsplitter. Similar to the power-recycling technique, a partially transmissive signal-recycling mirror (MSR) is installed to further resonantly enhance the GW-induced signal at the interferometer's output.}
BS: 50/50 beamsplitter, SHG: second harmonic generator, OPA: optical parametric amplifier, DBS: dichroic beamsplitter, PLL: phase locking loop, MFe/MFn: far interferometer end mirrors (east/north), MCe/MCn : central interferometer mirrors, T: mirror transmissivity. All interferometer optics are suspended by multi-stage pendulums and situated in a vacuum system.}
  \label{GEOsetup}
\end{figure*}

To date no direct detections of gravitational waves have been made. However, upper limits on GW signal strength for certain classes of sources could be derived, see \cite{LIGO-Nat09} and references there in. 
To realize gravitational wave astronomy with a daily event rate, a sensitivity improvement by at least one order of magnitude is required, see for example Ref.~\cite{PunturoETAL10}, thereby increasing the sensitive sky volume by 3 orders of magnitude. Science teams all over the world are currently addressing each instrumental noise source for future observatory generations. 
%
%
At higher audio-band frequencies it is the quantum nature of light that inhibits a more precise measurement, because the counting statistics of the light particles themselves lead to a fluctuating interferometer output (shot-noise). This noise is caused by so-called ``vacuum'', or ``zero-point'' fluctuations of the electro-magnetic field~\cite{Cav81}. The ``classical" approach to improve the observatory's signal-to-shot-noise ratio is an increase of the circulating light power, since the signals produced by gravitational waves are proportional to the light power, whereas the shot-noise is proportional to only the square root of the power. 
%
%
However, a higher light power leads to a thermal deformation of the sensitive interferometer optics and an increasing radiation pressure noise level resulting in a practical upper limit for the optical light power applicable \cite{PunturoETAL10}. Hence, additional technologies must be considered to push the sensitivity beyond this limitation \cite{SMML10}.

Squeezed states of light \cite{Wal83} provide a way of increasing the sensitivity in the shot-noise limited region, \textit{independently} of the circulating light power. Generally, a light field is described by two non-commuting physical quantities, the amplitude and phase quadratures. 
{The minimum product of their uncertainties is limited by Heisenberg's uncertainty relation (HUR), which is also valid in the complete absence of photons, i.e.~for a vacuum state. Vacuum states as well as coherent states have noise equally distributed in the field quadratures. Only squeezed states -- containing quantum correlated photons -- show a noise below the vacuum noise level, however, due to the HUR, this is not possible for all quadratures of the state simultaneously. For a Michelson interferometer operated close to a dark output port, squeezed states can be utilized by injecting them into the observatory's signal output port and spatially overlapping them with the high power laser field at the beam splitter~\cite{Cav81}. The squeezed quadrature has to be controlled such that, after being reflected off the interferometer, it is in phase with the readout (amplitude) quadrature of the observatory output light.} 
This scheme produces path entanglement between the high-power light fields in the interferometer arms and reduces the photon counting noise on the photo diode in a way that can be explained only by photon correlations that are stronger than any classical correlation~\cite{SMML10}. 

Since the first observation of squeezed light in 1985~\cite{SHYMV85}, squeezed light sources have constantly been improved, recently reaching a factor of almost 13\,dB below shot noise power~\cite{Eetal10}. At frequencies in the GW detection band, the generation of squeezing remained an unsolved problem for a long time. Only recently, squeezing at Fourier-frequencies in the audio-band~\cite{Metal04,VCDS07} and a coherent phase control scheme for squeezed vacuum states could be demonstrated~\cite{VCHFDS06}. In parallel, proof-of-principle experiments at higher frequencies have shown that small-scale Michelson-interferometer sensitivities can indeed be improved by squeezing~\cite{MSMBL02,Getal08}. Even though squeezed states are an ingredient for a multiplicity of quantum techniques like quantum teleportation~\cite{YAF04}, quantum memories~\cite{Polzik10} and many more, all of them are yet to mature from a proof-of-principle stage into a practical application. The increase of the GEO\,600 sensitivity below its shot-noise limit by non-classical means is indeed the first practical application of this quantum technology, with the potential to become an integral part for all future generations of laser-interferometric gravitational wave observatories.

\begin{figure}
\centerline{\includegraphics[width=7cm]{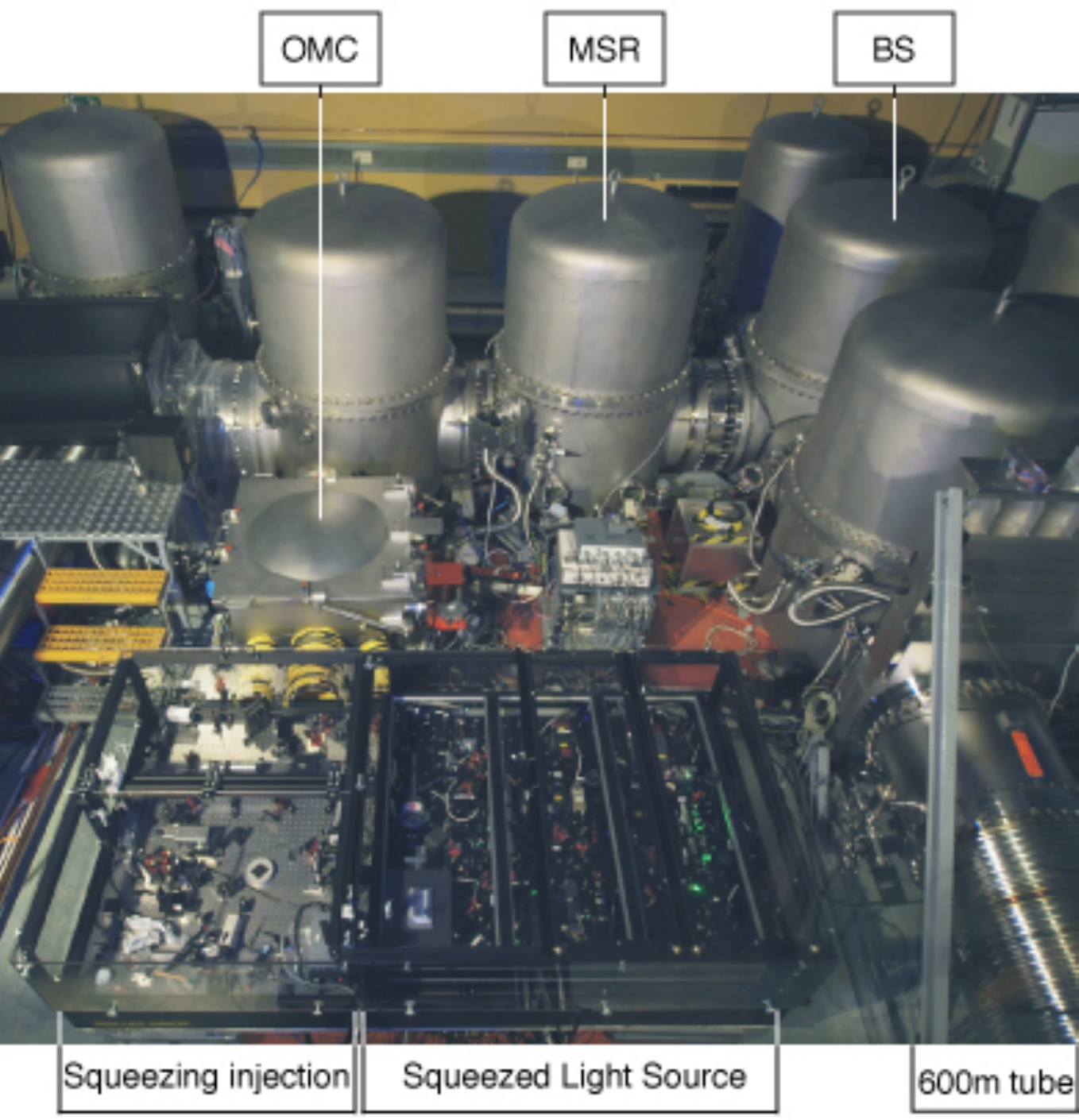}}
  \vspace{0mm}
\caption{View into the GEO\,600 central building. In the front, the squeezing bench containing the squeezed light source and the squeezing injection path is shown. The optical table is surrounded by several vacuum chambers containing suspended interferometer optics.}
  \label{GeoBild3}
\end{figure}

\begin{figure*}[t!]
\centerline{\includegraphics[width=\linewidth]{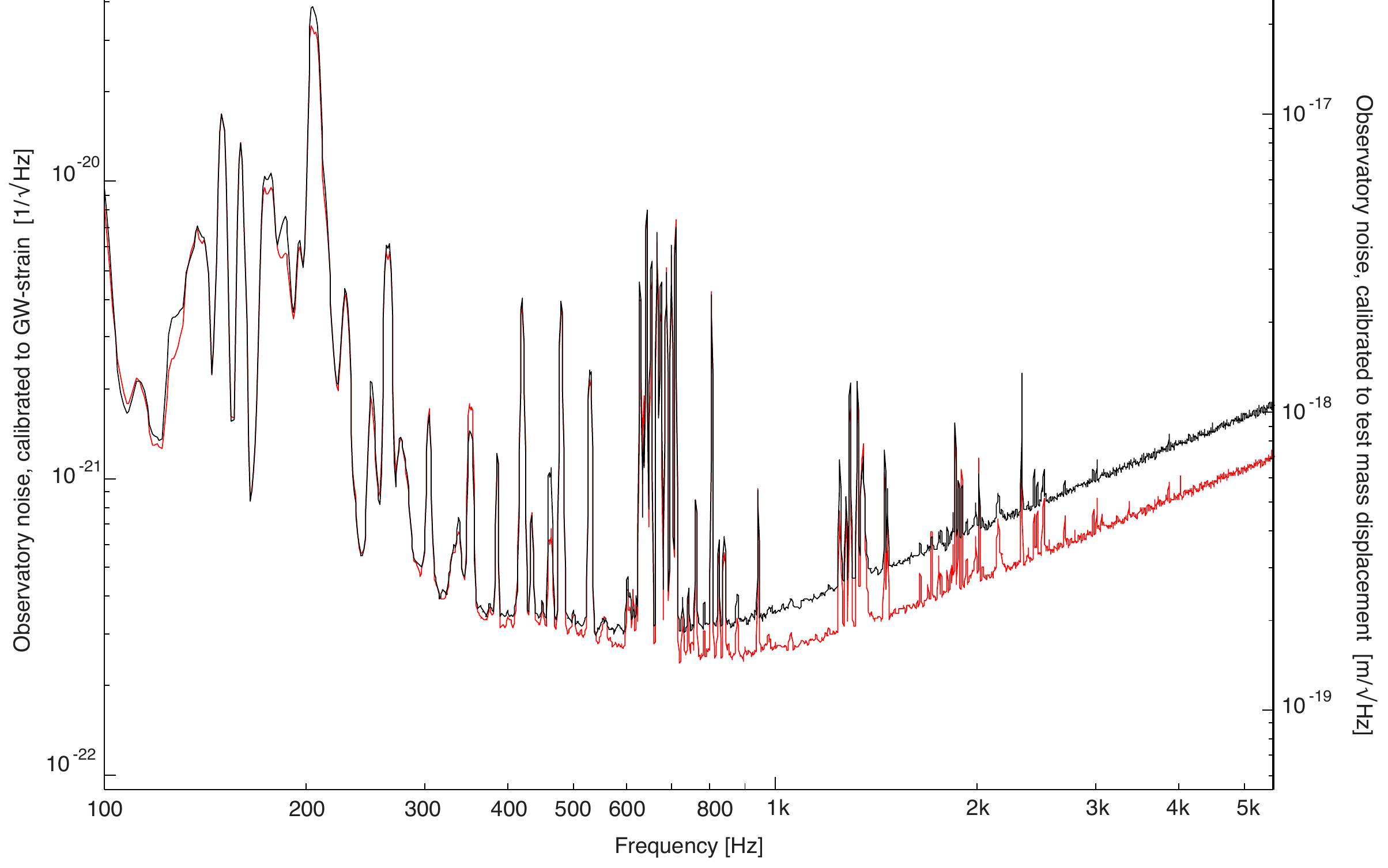}}
  \vspace{0mm}
\caption{Nonclassical reduction of the GEO\,600 instrumental noise using squeezed vacuum states of light, calibrated to GW-strain {and differential mirror displacement, respectively}. In black the observatory {noise spectral density} is shown without the injection of squeezed light. {Below 700\,Hz residual noise from the environment couples into the observatory. At higher frequencies GEO\,600 is shot-noise limited; note that the slope in the kHz-regime is due to the normalization and the frequency dependent signal enhancement of GEO\,600.} An injection of squeezed vacuum states into the interferometer leads to a broadband noise reduction of up to 3.5\,dB (red trace) in the shot-noise limited frequency band. The spectral features are caused by excited violin modes of the suspensions (600\,-\,700Hz and harmonics) as well as by calibration (160\,Hz\,-\,2.5\,kHz) and OMC alignment control (250\,-\,550\,Hz) lines.
Both traces were averaged over 4 minutes. The resolution bandwidth is 1\,Hz for frequencies below 1\,kHz, and 2\,Hz at higher frequencies.}
\label{GEOstrain}
\end{figure*}

The German-British GEO\,600 facility is one of the large-scale Michelson interferometers searching for gravitational wave signals. GEO\,600 already uses a number of so-called \emph{advanced techniques}, which are foreseen to be implemented in future upgrades of LIGO or Virgo \cite{Gretal08}. Fig.~\ref{GEOsetup} shows a simplified layout of GEO\,600. 
{The first steps of the ongoing GEO high frequency (HF) upgrade program have already been included~\cite{Luetal10}:} GEO\,600 is now operated in a tuned (to resonance with the laser carrier) signal-recycling mode with an optical homodyne detection scheme, also called \textit{DC readout}~\cite{Deetal10,Hietal09}; and the carrier light transmitted by the signal-recycling mirror is spatially and temporally filtered with an output mode cleaner cavity (OMC)~\cite{Deetal10}. 
The lower left part of Figure~\ref{GEOsetup} shows a simplified schematic of the squeezed light source which has been added in this work. The squeezed light beam is phase locked to the 12\,W GEO\,600 laser and comprises squeezed vacuum states at frequencies from 10\,Hz to above 10\,kHz and a MHz control field for stabilization of the squeezed quadrature with respect to the GW signal~\cite{VKLGDS10}. More details are given in the method summary.  
A view into the GEO\,600 central building, showing the squeezed light source and the parts of the vacuum system, is presented in Figure \ref{GeoBild3}.

Figure~\ref{GEOstrain} presents the result achieved by this work: the quantum technology enhancement of an operating gravitational wave observatory. The injection of squeezed vacuum states into GEO\,600 leads to a broadband noise reduction of up to 3.5\,dB (black to red trace) in the shot-noise limited frequency band (above 700\,Hz). The quantum noise at 3\,kHz was reduced from $1.0\times10^{-21}$ Hz$^{-1/2}$ down to $6.7\times10^{-22}$ Hz$^{-1/2}$. This corresponds to a factor 1.5$^3$ $\approx$ 3.4 increase in detection rate for isotropically distributed GW sources in that frequency band. 
{The squeezing enhancement has been successfully operated for several consecutive hours just limited by the current performance of the beam alignment.}
Due to the application of squeezed light the GW observatory GEO600 {has now achieved} its best ever sensitivity since the implementation of the advanced homodyne detection scheme. As expected, at Fourier frequencies below 700\,Hz, squeezed light does neither reduce nor increase the current displacement noise level of about $10^{-18}\,\rm{m}/\sqrt{\rm{Hz}}$. This observation makes us confident that a squeezed light improvement will extend to these frequencies as soon as the currently limiting technical noise will be reduced. Note, that quantum radiation pressure noise \cite{Cav80} is not expected to be significant at these frequencies at the present sensitivity.

The measured nonclassical quantum noise reduction in GEO\,600 presented here is not limited by the squeezed light laser but by optical loss on the squeezed light during propagation in the interferometer. The 10\,dB injected squeezed state is degraded by photon absorption and scattering inside the GEO\,600 signal recycling cavity and the output mode-cleaner, both contributing about 10\,\% loss. In addition, the non-perfect photo diode quantum-efficiency, the absorption of the Faraday isolators and auxiliary optics, and finally some residual mode mismatch cause an additional 20\,\% loss. All losses have been verified by independent measurements and provide an overall optical efficiency of $\eta=0.62$. This leads to a reduction of the nonclassical noise suppression (in power) from $V_{\rm{sqz}}=0.1$ (or $10$\,dB) to $\eta V_{\rm{sqz}} +(1-\eta)=0.44\,($or 3.5\,dB), in excellent agreement with our results shown in Figure~\ref{GEOstrain}. Based on this, we are confident that  future optical loss reductions will result in a correspondingly higher squeezing factor. During the GEO-HF upgrade program in 2011 we expect a sensitivity improvement of up to 6\,dB to be realized with squeezed light input.
An even stronger impact through the application of squeezed light can be foreseen in future gravitational wave observatories, where state of the art optical technologies will allow for lower optical losses.

The results presented here show that squeezed light can improve operating gravitational wave observatories. 
Since squeezed light is also highly compatible with thermal noise reduction by means of cryogenic cooling of observatories \cite{SMML10}, we expect this innovative approach to become a key technology in making gravitational wave astronomy a reality, and we 
believe that squeezed light lasers in addition to high-power lasers are likely to be integrated into all future gravitational wave observatories.

\subsection*{Method summary}
Altogether four different laser frequencies are involved in the generation and coherent control of the squeezed vacuum states, see fig.~\ref{GEOsetup}. The main 2\,W laser, which is phase locked to the 12\,W GEO\,600 laser, drives a second-harmonic generator (SHG). The green light from the SHG is filtered using a ring-resonator to attenuate high-frequency phase noise~\cite{FHDFS06}. The frequency up-converted field is subsequently injected into the squeezing resonator containing  a nonlinear medium (periodically poled potassium titanyl phosphate) placed in a standing-wave half-monolithic cavity. Only 35\,mW of the frequency doubled field are required to generate {about 9\,dB} squeezing down to 10\,Hz~\cite{VKLGDS10}, via the process of parametric down-conversion  (optical parametric amplification). In order to avoid any contamination of the squeezed light by laser noise in the audio band, two auxiliary lasers, frequency-shifted by several MHz, are employed for coherent control of the squeezed vacuum states~\cite{VCHFDS06,VKLGDS10}. Squeezing at Fourier-frequencies in the audio-band has been shown to be very sensitive to light back-scattered into the OPA~\cite{Metal04, VCHFDS06,VCDS07}. Therefore, the squeezed beam is guided through two Faraday isolator units before it is injected into the signal port of GEO\,600.\\~\\~\\

%
\newpage
\section*{Acknowledgement}
The authors gratefully acknowledge the support of the United States
National Science Foundation for the construction and operation of the
LIGO Laboratory and the Science and Technology Facilities Council of the
United Kingdom, the Max Planck Society, the Deutsche Forschungsgemeinschaft, the cluster of excellence QUEST (Centre for Quantum Engineering and Space-Time Research),  the BMBF, the Volkswagen Foundation, and the State of
Niedersachsen/Germany for support of the construction and operation of
the GEO\,600 detector. The authors also gratefully acknowledge the support
of the research by these agencies and by the international Max Planck Research School (IMPRS), the SFB\,TR7, the FP7 project Q-ESSENCE, the Australian Research Council,
the Council of Scientific and Industrial Research of India, the Istituto
Nazionale di Fisica Nucleare of Italy, the Spanish Ministerio de
Educaci\'on y Ciencia, the Conselleria d'Economia, Hisenda i Innovaci\'o of
the Govern de les Illes Balears, the Royal Society, the Scottish Funding 
Council, the Scottish Universities Physics Alliance, The National Aeronautics 
and Space Administration, the Carnegie Trust, the Leverhulme Trust, the David
and Lucile Packard Foundation, the Research Corporation, and the Alfred
P. Sloan Foundation.



%

\begin{thebibliography}{10}

\bibitem{Ein16} Einstein, A. Die Grundlage der allgemeinen Rela\-tivit\"atstheorie, \textit{Annalen der Physik} \textbf{49}, 769-822 (1916).

\bibitem{SSc09} Sathyaprakash, B.~S.~\& Schutz, B.~F. Physics, Astrophysics and Cosmology with Gravitational Waves. \textit{Living Rev. Relativity} \textbf{12}, 2 (2009).

\bibitem{Cav81} Caves, C.~M. Quantum-mechanical noise in an interferometer. \textit{Phys. Rev. D} \textbf{23}, 1693-1708 (1981).

 \bibitem{Wei72} Weiss, R. Electromagnetically Coupled Broadband Gravitational Antenna. Quarterly Progress Report, Research Laboratory of Electronics, MIT 105: 54 (1972).

\bibitem{AbbottETAL09ligo} Abbott, B.~P.~\textit{et al.} LIGO: the Laser Interferometer Gravitational-Wave Observatory. \textit{Rep. Prog. Phys.} \textbf{72}, 076901 (2009).

\bibitem{Acernese08} Acernese, F.~\textit{et al.} Status of Virgo. \textit{Class. Quantum Grav.} \textbf{25}, 114045 (2008).

\bibitem{Wetal02} Willke, B.~\textit{et al.} The GEO\,600 gravitational wave detector. \textit{Class. Quantum Grav.} \textbf{19}, 1377-1387 (2002).

\bibitem{Gretal10} Grote, H.~\textit{et al.} The GEO\,600 status. \textit{Class. Quantum Grav.} \textbf{27}, 084003 (2010).

\bibitem{AraiETAL09} Arai, K.~\textit{et al.} Status of Japanese gravitational wave detectors. \textit{Class. Quantum Grav.} \textbf{26}, 204020 (2009).

\bibitem{BarrigaETAL10} Barriga, P.~\textit{et al.} AIGO: a southern hemisphere detector for the worldwide array of ground-based interferometric gravitational wave detectors. \textit{Class. Quantum Grav.} \textbf{27}, 084005 (2010).

\bibitem{PunturoETAL10} Punturo, M.~\textit{et al.} The third generation of gravitational wave observatories and their science reach. \textit{Class. Quantum Grav.} \textbf{27}, 084007 (2010).

\bibitem{LIGO-Nat09} The LIGO Scientific Collaboration \& The Virgo Collaboration. An upper limit on the stochastic gravitational-wave background of cosmological origin. \textit{Nature} \textbf{460}, 990-994 (2009).

\bibitem{SMML10} Schnabel, R., Mavalvala, N., McClelland, D.~E.~\&~Lam, P.~K. Quantum metrology for gravitational wave astronomy. \textit{Nat. Commun.} \textbf{1:121} doi: 10.1038/ncomms1122 (2010).

\bibitem{Wal83} Walls, D.~F. Squeezed states of light. \textit{Nature} \textbf{306}, 141-146 (1983).

\bibitem{SHYMV85} Slusher, R.~E., Hollberg, L.~W., Yurke, B., Mertz, J.~C.~\&~Valley, J.~F. Observation of Squeezed States Generated by Four-Wave Mixing in an Optical Cavity. \textit{Phys. Rev. Lett.} \textbf{55}, 2409-2412 (1985).

\bibitem{Eetal10} Eberle, T.~\textit{et al.} Quantum Enhancement of the Zero-Area Sagnac Interferometer Topology for Gravitational Wave Detection. \textit{Phys. Rev. Lett.} \textbf{104}, 251102 (2010).

\bibitem{Metal04} McKenzie, K.~\textit{et al.} Squeezing in the Audio Gravitational-Wave Detection Band. \textit{Phys. Rev. Lett.} \textbf{93}, 161105 (2004).

\bibitem{VCDS07} Vahlbruch, H., Chelkowski, S., Danzmann, K.~\&~Schnabel, R. Quantum engineering of squeezed states for quantum communication and metrology. \textit{New J. Phys.} \textbf{9}, 371 (2007).

\bibitem{VCHFDS06} Vahlbruch, H., Chelkowski, S., Hage, B., Franzen, A., Danzmann, K.~\&~Schnabel, R. Coherent Control of Vacuum Squeezing in the Gravitational-Wave Detection Band. \textit{Phys. Rev. Lett.} \textbf{97}, 011101 (2006).

\bibitem{MSMBL02} McKenzie, K., Shaddock, D.~A., McClelland, D.~E., Buchler, B.~C.~\&~Lam, P.~K. Experimental Demonstration of a Squeezing-Enhanced Power-Recycled Michelson Interferometer for Gravitational Wave Detection. \textit{Phys. Rev. Lett.} \textbf{88}, 231102 (2002).

\bibitem{Getal08} Goda, K.~\textit{et al.} A quantum-enhanced prototype gravitational-wave detector. \textit{Nature Phys.} \textbf{4}, 472-476 (2008).


\bibitem{YAF04} Yonezawa, H., Aoki, T.~\&~Furusawa, A.  Demonstration of a quantum teleportation network for continuous variables. \textit{Nature} \textbf{431}, 430-433 (2004).


\bibitem{Polzik10} Jensen, K.~\textit{et al.} 
Quantum memory for entangled continuous-variable states. \textit{Nature Phys.} \textbf{7}, 13Ð16 (2010).

\bibitem{Gretal08} Grote, H.~\textit{et al.} The GEO\,600 status. \textit{Class. Quantum Grav.} \textbf{25}, 114030 (2008).


\bibitem{Luetal10} L\"uck, H.~\textit{et al.} The upgrade of GEO\,600. \textit{J. Phys. Conf. Ser.} \textbf{228}, 012012 (2010).

\bibitem{Deetal10} Degallaix, J.~\textit{et al.} Commissioning of the tuned DC readout at GEO\,600 \textit{J. Phys. Conf. Ser.} \textbf{228}, 012013 (2010).

\bibitem{Hietal09} Hild, S.~\textit{et al.} DC-readout of a signal-recycled gravitational wave detector. \textit{Class. Quantum Grav.} \textbf{26}, 055012 (2009).

\bibitem{VKLGDS10} Vahlbruch, H., Khalaidovski, A., Lastzka, N., Gr{\"a}f, C., Danzmann, K.~\&~Schnabel, R. The GEO600 squeezed light source.
{\em Class. Quantum Grav.} {\bf 27}, 084027 (2010).

\bibitem{Cav80} Caves, C.~M. Quantum-Mechanical Radiation-Pressure Fluctuations in an Interferometer. \textit{Phys. Rev. Lett.} \textbf{45}, 75--79 (1980).

\bibitem{FHDFS06} Franzen, A., Hage, B., DiGuglielmo, J., Fiur\'{a}\v{s}ek, J.~\&~Schnabel, R. Experimental demonstration of continuous variable purification of squeezed states. {\em Phys. Rev. Lett.} \textbf{97}, 150505 (2006).



\end{thebibliography}
\end{document}